\documentclass[%
 reprint,
superscriptaddress,
longbibliography,
 amsmath,amssymb,
 aps,
]{revtex4-2}

\usepackage{graphicx}
\usepackage{dcolumn}
\usepackage{bm}
\usepackage{multirow}

\begin{document}

\preprint{APS/123-QED}

\title{Probing wetting and solidification dynamics of neon thin films with a high-$T_c$ superconducting circuit}

\author{Brennan Dizdar}
\thanks{These authors contributed equally to this work.}
\affiliation{James Franck Institute and Department of Physics, University of Chicago, Chicago, Illinois, 60637, USA}

\author{Yueheng Shi}
\thanks{These authors contributed equally to this work.}
\affiliation{Departments of Physics and Applied Physics, Stanford University, Stanford, California, 94305, USA}

\author{David I. Schuster}
\affiliation{Departments of Physics and Applied Physics, Stanford University, Stanford, California, 94305, USA}

\author{Christopher S. Lang}
 \email{christopher.lang@anl.gov}
\affiliation{James Franck Institute and Department of Physics, University of Chicago, Chicago, Illinois, 60637, USA}
\affiliation{Q-NEXT, Argonne National Laboratory, Lemont, Illinois, 60439, USA}

\date{\today}

\begin{abstract}

Integrating condensed noble gases with superconducting circuits offers new opportunities for operating hybrid quantum devices. 
In particular, electrons on solid neon have recently emerged as exceptionally coherent charge qubits, however the growth of the underlying neon films on patterned devices remains poorly controlled.
We monitor neon films near the triple point using a multimode resonator fabricated from high-$T_c$ YBCO, which remains superconducting at temperatures where the liquid phase exists and whose geometry sets a thickness-sensing window near 1 $\mu$m.  
In the liquid state we observe rich film dynamics: uniform wetting that saturates near 20 nm, drainage of thicker films, and spontaneous thickness oscillations, reflecting the extreme sensitivity of thick films in this regime.
A galvanic connection to our sensing volume enables in-situ electrostatic tuning of the liquid film thickness.
Statistics of freezing the neon across various liquid levels reveal variations of the solid film thickness by approximately an order of magnitude.
We also observe a two-fold increase in the quality factor of the resonators at the level of $10^3$ with added neon and rule out dielectric loss as the limiting factor.
Our results demonstrate that hermetic packaging facilitates the investigation of both rich non-equilibrium fluid dynamics and the impact of condensed noble gas films on microwave loss in superconducting circuits, but may not be the best modality for growing solid neon films with precisely controlled thickness.

\end{abstract}

\maketitle


\section{\label{sec:intro} Introduction}

Understanding how superconducting circuits couple to external degrees of freedom is at the heart of both developing hybrid quantum architectures and mitigating dielectric loss \cite{clerk_hybrid_2020}. 
Condensed noble gases are an attractive ingredient for such architectures: they are chemically pure, nearly free of nuclear spins, can be grown and tuned in situ at cryogenic temperatures, and support pristine two-dimensional electron gases at their surfaces, motivating proposals for scalable, reconfigurable qubit arrays \cite{lyon_spin-based_2006,kawakami_blueprint_2023}.
In particular, solid neon has recently emerged as a host for both high-coherence electron spin ensembles and single-electron charge qubits with coherence times on the order of $\sim$100 $\mu$s \cite{kanagin_impurities_2025, zhou_electron_2024}.
Much of this coherence derives from the rigidity of the solid surface: unlike electrons on liquid helium, whose motional linewidths are limited at the $\sim$100 MHz level by coupling to surface capillary waves (ripplons) \cite{koolstra_strong_2026}, electrons on neon are immune to this decoherence channel.
Despite this progress, the precise nature of the electron wavefunction and the impact of the vacuum/neon/substrate interfaces on qubit coherence are not fully understood \cite{zheng_surface-morphology-assisted_2025}.
A prerequisite for investigating these questions is understanding the underlying solid neon film itself and how its growth conditions determine its properties.

Solid neon films can be grown in two ways: 1) quench-condensation of gaseous neon directly onto a cold substrate held at a temperature below the triple point or 2) cooling a substrate with a liquid neon film through the triple point.
It is known that for liquid neon adiabatically cooled through the triple point $T_t = 24.56$K, the resulting solid film thickness saturates to just a few monolayers as dictated by triple point wetting physics \cite{migone_triple-point_1986}.
Experiments that have reported charge qubits on solid neon have exclusively used the latter method to freeze a liquid film to obtain a solid film thicker than a few monolayers; such a scenario suggests a non-equilibrium freezing modality that warrants further investigation.
One challenge in efficiently characterizing the neon film growth is that the triple point of neon is much higher than the temperatures typically suitable for superconducting measurements.
Indeed, electron-on-neon charge qubit experiments with conventional superconducting resonators are unable to detect properties of the liquid state when neon is introduced into the hermetic sample cell \cite{zhou_single_2022,zhou_electron_2024,li_solid_2026}.
This challenge can be circumvented by using high $T_c$ superconducting resonators \cite{duthaluru_real-time_2026}.

\begin{figure*}
\includegraphics{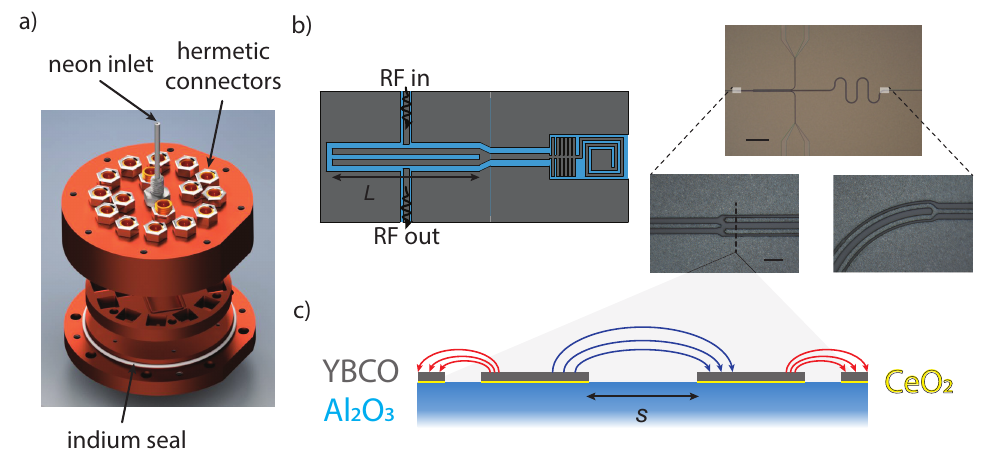}
\caption{\label{fig:fig1} (a) Rendering of the hermetic sample cell used in the experiment \cite{koolstra_trapping_2019}. (b) (Left) Schematic of our DC-coupled coplanar stripline resonator. The differential mode frequency is determined by the coplanar strip length $L$ where the voltage node is located where the pins join. An on-chip LC low pass filter, consisting of a set of interdigitated finger capacitors in series with a spiral inductor to a bond pad, helps suppress leakage of the common mode. (Right) Optical image of the device, with zoom-ins of the voltage anti-node of the coplanar stripline (left) and where the two pins join to define the voltage node for the differential mode (right). Scale bars correspond to 50 $\mu$m (top) and 10 $\mu$m (bottom). The DC gate on the left side of the device is unused in this experiment. (c) Cross section of the materials stack. The sample is fabricated from a 300 nm thick film of YBCO deposited on a sapphire substrate with an intermediate buffer layer of 40 nm thick $\textrm{CeO}_2$ in between. Electric field lines of the differential (blue) and common (red) modes are shown. The vertical sensing region is determined by the gap width $s$.}
\end{figure*}

In this work, we operate a voltage tunable multimode superconducting resonator fabricated from a high-$T_c$ YBCO film to sense the dynamics of liquid and solid neon films in real time.
We identify qualitatively different regimes of liquid film dynamics ranging from uniform wetting to spontaneous oscillations, all of which have implications for thickness control.
Repeated solidification trials via cooling through the triple point reveal only a weak correlation between initial liquid thickness and final solid thickness and an overall thinning of the film.
Finally, we observe a smooth increase in the quality factor of our resonators as a function of neon thickness; such a trend suggests a potential intricate interplay between the neon film and quasiparticle loss.

\section{\label{sec:setup} Experimental setup}

Our measurement is based on the sensitivity of a planar microwave resonator to its dielectric environment.
The electric field of a coplanar stripline mode extends above the chip surface over a distance set by the transverse dimensions of the stripline.
When neon condenses onto the chip, it replaces vacuum ($\varepsilon_r = 1$) with a dielectric (liquid $\varepsilon_r \approx 1.19$, solid $\varepsilon_r \approx 1.24$ \cite{cole_image-potential-induced_1969}) within this sensing volume, increasing the capacitance per unit length and lowering the resonance frequency.
Continuously tracking the resonance frequency therefore provides an in-situ, real-time record of the neon film thickness on the resonator, with a thickness window set by the field decay length.
Because the temperature range of interest lies far above the transition temperatures of conventional superconductors, we fabricate the resonator from the high-$T_c$ superconductor YBCO ($T_c \approx 88$ K), which remains deep in its superconducting state at the neon triple point ($T_t = 24.56$K).

In our experiment, we monitor two modes of a YBCO-on-sapphire coplanar stripline resonator (fabricated by STAR Cryoelectronics) enclosed in a hermetic sample cell (Fig. \ref{fig:fig1}a).
In this work, the cell is mounted to the mixing chamber of a standard Bluefors LD250 dilution refrigerator.
The cell is cooled by the pulse tube through the two sets of heat switches between the 4K and mixing chamber stages.
The cooling rate of the cell is controlled by adjusting the power applied to a resistive heater mounted at the 4K stage; this enables stabilization of the mixing chamber at a desired temperature around the triple point.

The resonator, measured in a transmission configuration, is galvanically connected to a DC voltage source that enables control over the local electric field between the center pins and ground (Fig. \ref{fig:fig1}b).
The length of the section before and after the two center pins are joined determining the frequencies of the common and differential modes (Fig. \ref{fig:fig1}b).
For the results presented in this paper, we measure the only two modes that are present in our measurement band: the fundamental $\lambda/4$-differential and the first overtone $3\lambda/4$-common mode.
An independent measurement in a separate cooldown utilizing a 180-degree hybrid confirms the common/differential nature of these modes and shows that they are in the undercoupled regime $Q_\textrm{tot} \approx Q_\textrm{int}$ (see Appendix \ref{app:dc_rf_characterization}).

\begin{figure*}
\includegraphics{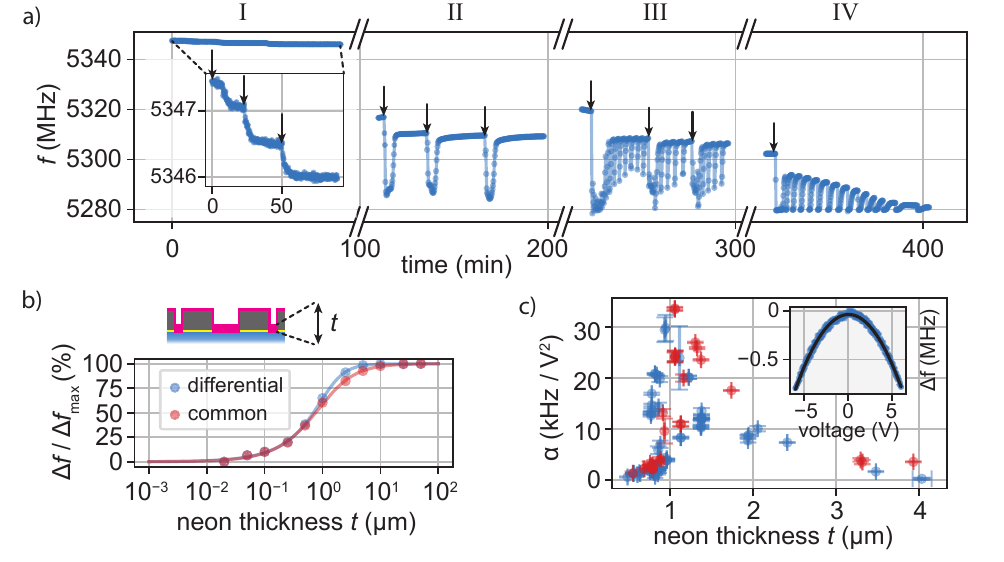}%
\caption{\label{fig:fig2} (a) Monitoring the differential mode resonance frequency as liquid neon is filled into the cell held at $T = 26$ K. Arrows indicate discrete filling events. Data segments for various qualitative regimes (I - IV) are spliced together on same plot with axes breaks to show relative timing. (b) Simulated finite element model of estimating the neon thickness $t$ from the frequency shift of the resonator modes. (c) In-situ dielectric loading tunability via an applied DC voltage for liquid films of various thicknesses. Each individual trace (example shown in inset) displays no hysteresis.}
\end{figure*}

\section{Liquid film dynamics}
We first characterize condensation in the liquid state by periodically admitting metered doses of neon gas into the sample cell held at $T=26$K, above the triple point.
Neon enters the fill line via a gas manifold which operates by pressurizing a 10 cm$^3$ control volume with gaseous neon to a pressure above the triple point pressure ($P_t \approx$ 433 mbar) and then opening a switch to the cryostat. A detailed description of the manifold can be found in Ref. \cite{koolstra_trapping_2019}.
In our experiment, pressures between 1 and 3 bar were used. Each dose dielectrically loads the mode volumes and shifts the resonance frequencies downward as shown in Fig. \ref{fig:fig2}a for the differential mode.
We observe four qualitatively different regimes.
In regime I the frequency decreases gradually and monotonically with each dose, consistent with gradual uniform wetting of the entire surface with a thin liquid film.
This film saturates to a thickness of approximately 20 nm (inferred from the model described below), after which regime II begins.
Each insertion displays a large initial shift that relaxes toward a steady state, suggesting that each insertion covers the resonator with a thick liquid film that subsequently thins as neon redistributes to the rest of the cell.
In regime III, the liquid level becomes unstable, exhibiting fluctuations in level as periodic oscillations occur even without neon insertions.
This susceptibility is plausibly related to the enhanced evaporative response of a thin film near saturation, where the rate of exchange between liquid and vapor on the surface of the film sets the dynamics of the film \cite{OronDavisBankoff1997,CrasterMatar2009}.
Such liquid thin films are known to develop periodic thickness oscillations once the heat flux from the substrate exceeds a threshold \cite{Nguyen2018}.
Finally in regime IV the frequency shift saturates to a maximum observed value which we identify as the resonator mode volumes being completely filled with neon.

To translate the observed frequency shift to an effective neon thickness $t$, we utilize a model obtained from finite-element simulations (Ansys Maxwell) of the cross sectional mode capacitances as the vacuum above the resonator pins is progressively replaced with neon of various heights (Fig. \ref{fig:fig2}b) (see Appendix \ref{app:dielectric_filling_model}).
This provides a mapping between neon thickness and frequency shift $f_\textrm{meas} = f_0 + \delta f(t)$ which we are able to invert to obtain $t(f_\textrm{meas})$ under the assumption that the neon film is uniform across the surface of the resonator.
The resulting calibration curve highlights a key design consideration:  the transverse dimensions of the resonator determine the neon height $t^*$ that maximizes the sensitivity $df/dt$, which in our work is around $t^* \approx$ 1 $\mu\textrm{m}$.
This geometric tunability suggests that resonators with different cross sections could be multiplexed on a single chip to extend the high-precision window over a larger dynamic range of film thicknesses.

A defining characteristic of a liquid thin film covering the resonator is the ability to in-situ tune the film thickness via a local electric field, as has been shown for superfluid helium \cite{yang_coupling_2016}.
Applying a DC voltage $V$ to the center pins produces an inhomogeneous field that polarizes the liquid and draws it toward the resonator, thickening the film and shifting the resonance frequency by $\delta f = \alpha V^2$.  We demonstrate this tunability by sweeping the DC voltage $V$ of the resonator center pins across inferred liquid film thicknesses, spanning the sensitivity window.
The quadratic coefficients of the resulting frequency shifts are plotted in Fig. \ref{fig:fig2}c. 
Individual voltage sweeps show no hysteresis indicating the film responds reversibly and is free of pinning.  However, the variance in the data for different trials with similar estimated thicknesses suggests that the liquid is not completely uniform across the resonator's surface.

\begin{figure}[b]
\includegraphics{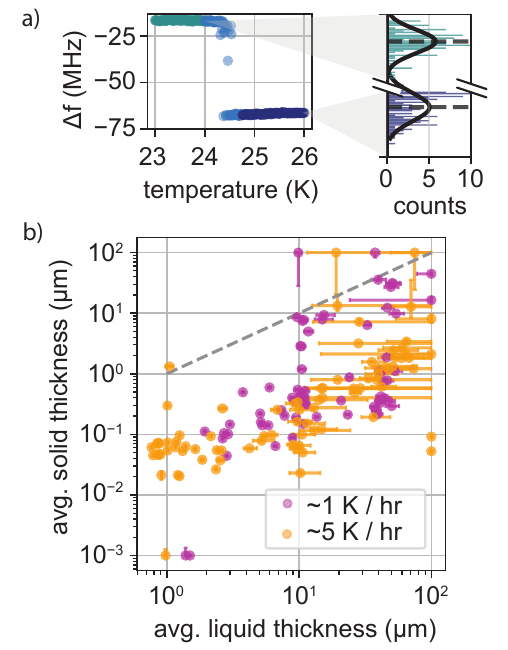}
\caption{\label{fig:fig3} (a) Left: representative trace of the frequency shift of the differential mode as the mixing chamber is cooled through the freezing point. Right: the frequency data is binned in the temperature ranges $T \in [23,24.2]$ K and $T \in [24.75, 26]$ K for the solid and liquid thickness estimation, respectively. (b) The average film thickness in the liquid and solid states determined from the histogram fits in (a) are plotted against each other across many different trials at two difference cooling rates. Within our statistics, the two cooling rates yield similar distributions. Gray dashed line indicates equal thickness in the liquid and solid states.}
\end{figure}

\section{Neon solidification}
We next investigate solidification by cooling the cell from $T=26$K through the triple point.
Fig. \ref{fig:fig3}a shows a representative trace tracking the frequency of the differential mode as the system is cooled; at the freezing transition the magnitude of the frequency shift abruptly decreases, indicating the film almost always becomes thinner upon solidification.
To quantify each trial, we bin the measured frequencies away from the transition ($T\in [24.75,26]$K) for the liquid and ($T\in [23,24.2]$K) for the solid and  convert to an average thickness using our finite-element model.

The central question of our work is whether control of the liquid film thickness (particularly in regimes I and II) can translate to a controlled solid film thickness.
Fig. \ref{fig:fig3}b shows the estimated average solid film thickness plotted against the corresponding average liquid film thickness for many repeated trials across the two different cooling rates of approximately $\sim$1K/hr and $\sim$5K/hr.
The solid films are typically one to two orders of magnitude thinner than the parent liquid, far thicker than a few monolayers but far thinner than the liquid they froze from, consistent with solidification proceeding in a non-equilibrium dewetting process. 
Thicker liquid films do on average yield thicker solid films, but the correlation is weak: for liquid films in the 1-10 $\mu$m range, where the thickness estimate is most precise the resulting solid thickness is reproducible only to within roughly an order of magnitude, while thicker liquid films (t $>$ 10 $\mu$m, for which sensitivity is also reduced) produce solid thicknesses spanning nearly the full observable range. 
The general spread of the distribution and lack of a strong influence of the different cooling rates is consistent with previous work \cite{duthaluru_real-time_2026}.

\begin{figure}[b]
\includegraphics{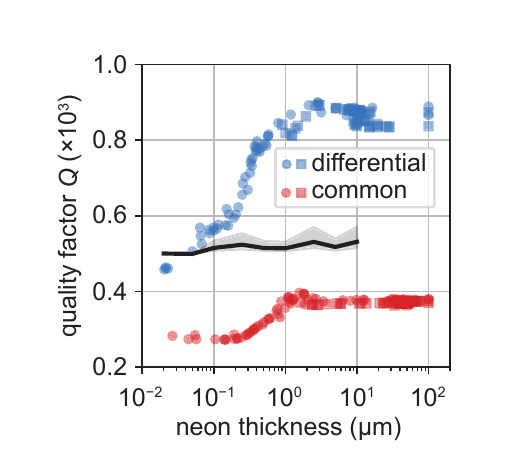}
\caption{\label{fig:fig4} Loaded quality factor $Q_L$ of the common (red) and differential (blue) modes as a function of neon level in both solid (circles, $T = 23.5$K) and liquid (squares, $T = 25$K) states. A dielectric loss model attributing all the loss to the decreasing participation ratio of the $\textrm{CeO}_2$ buffer layer is insufficient to explain the data (black line). The shaded region represents a factor of two range of the dielectric constant of $\textrm{CeO}_2$.}
\end{figure}

\section{Impact of neon on quality factor}
The presence of neon films on top of the resonator modifies the electromagnetic environment of the mode volumes; it is thus natural to investigate the impact of the neon film thickness on the quality factor of the modes.
Strikingly, we find that the quality factor smoothly increases (by up to $2\times$) with neon film thickness for both liquid and solid films (Fig. \ref{fig:fig4}).
It is natural to ask whether or not this increase can be explained by a dielectric loss model:
\begin{align}
    \frac{1}{Q(t)} =\frac{1}{Q_\textrm{other}} + \sum_i \frac{p_i(t)}{q_i}
\end{align}
where $p_i(t)$ are the thickness dependent participation ratios of the neon film, CeO$_2$ buffer layer, a surface region for both the YBCO-vacuum and sapphire-vacuum interfaces, and the bulk sapphire, and $q_i$ are the quality factors of each region.
As the neon film thickness increases, $p_\textrm{neon}(t)$ increases while all other participation ratios adjust accordingly.
Thus, an increase in the overall quality factor would suggest that $Q$ is limited by a lossy dielectric region whose participation decreases with the neon thickness; in this case, this would correspond to the CeO$_2$ buffer layer.
However, finite element simulations of the participation ratios rule out this scenario (see App.~\ref{app:participation_simulations}).
This conclusion is consistent with the fact that YBCO resonators are typically limited by conductor loss, especially at elevated temperatures \cite{gao_microwave_1993, velluire-pellat_hybrid_2023}.
We leave a detailed investigation of the interplay between the neon film and quasiparticle loss mechanisms as a subject for future work.
Our results motivate a deeper and continued investigation of how condensed noble gas films influence the quality factor of superconducting resonators.
Recent work on this front has demonstrated that incorporating condensed cryogenic films on superconducting circuits can reduce the absorption of stray terahertz radiation \cite{valenti_hydrogen_2024} or potentially remove surface impurities \cite{matkovic_characterizing_2025}.

\section{\label{sec:conclusion} Conclusion and outlook}

In this work, we utilized a voltage-tunable multimode coplanar stripline resonator in a hermetic cell to systematically explore dynamics across the entire range of liquid neon thicknesses in the resonator's modal sensing volume.
The data reveal rich non-equilibrium dynamics of the liquid film that likely contribute to reduced uniformity across the sample surface.
This may be due to the many thermal links between the cooling source and the substrate; more direct control of the local temperature may help facilitate better control of the film properties \cite{duthaluru_real-time_2026}.
Repeated solidification trials reveal that the solid film thickness is set only coarsely by the liquid film from which it froze, with an order of magnitude spread that is insensitive to the cooling rate over the range explored.
We also find that the neon film increases the quality factor of both resonator modes, an improvement that cannot be attributed to dielectric loading and that adds to growing evidence that condensed-gas overcoats can benefit superconducting circuits in certain circumstances.

Thus far, electron-on-neon qubits have only been demonstrated on solid thin films grown via freezing of a liquid film in a hermetic cell.
Our results motivate the investigation of such qubits on quench-condensed films which have better thickness control \cite{kanagin_impurities_2025}.
Such an approach is compatible with cryogenic atomic force microscopy, which would be a powerful complementary technique for obtaining spatial information on quantities of interest like surface roughness which cannot directly be resolved with resonator measurements.
This would allow a systematic study of the influence of both neon thickness and roughness on qubit properties, which are likely to be heavily influenced by surface disorder \cite{zheng_surface-morphology-assisted_2025}.
Ultimately, the ability to grow solid neon films with controllable thickness and low roughness would potentially enable the operation of a high mobility two-dimensional electron gas for reconfigurable electron spin qubits without suffering the loss channel of ripplons as is the case for liquid helium.

\begin{acknowledgments}
This work was primarily supported by the Air Force Office of Scientific Research (AFOSR) under Award no. FA9550-23-1-0636.
C.S.L. acknowledges support of both the Grainger Fellowship from the University of Chicago and the Department of Energy's Q-NEXT National Quantum Information Science Research Center.
\end{acknowledgments}

\appendix

\section{\label{app:dc_rf_characterization} Device characterization}

\begin{figure}[b]
\includegraphics{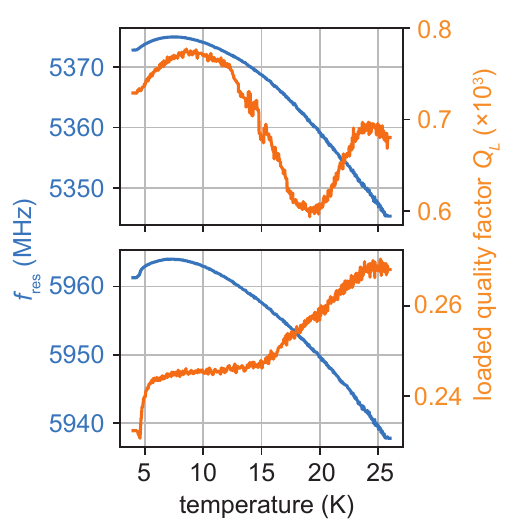}
\caption{\label{fig:sfig1} Temperature dependence of the resonant frequency $f_\textrm{res}$ and loaded quality factor $Q_L$ of the differential (top) and common (bottom) modes.}
\end{figure}

Here we describe the reference characterization of our sample.
A resistivity measurement of the YBCO film reveals a critical temperature $T_c \approx$ 88 K and a linear dependence of the resistivity $\rho = \rho_0 + \alpha (T - T_c)$ for $T>T_c$ with $\rho_0 \approx 50 \, \mu\Omega \cdot \textrm{cm}$ and $\alpha \approx 0.625 \, \mu\Omega \cdot \textrm{cm/K}$.

In the experiment, we only observe the two modes presented in the main text in our measurement band (4 - 8 GHz).
This is supported by HFSS simulations of the entire device which predicts mode frequencies of approximately 5.9 GHz and 6.1 GHz for the $\lambda/4$ differential and $3\lambda/4$ common modes, respectively.
The simulation does not capture the effects of kinetic inductance and the CeO$_2$ buffer layer, which likely contribute to the discrepancy between the simulated and measured values.
The simulation also reveals that a significant fraction of the common mode volume lives in the on-chip low pass filter, whereas this is not true for the differential mode as expected. 
The temperature dependence of the mode frequencies and quality factor are shown in Fig. \ref{fig:sfig1}.

\section{\label{app:dielectric_filling_model} Dielectric filling model}

\begin{table}[h]
\centering
\begin{tabular}{|c|c|c|}
\hline
 & Simulated & Measured \\
\hline
Differential, liquid & 66.96 & 67.99 \\
\hline
Differential, solid & 83.78 & 80.28 \\
\hline
Common, liquid & 73.53 & 70.92 \\
\hline
Common, solid & 92.00 & 82.13 \\
\hline
\end{tabular}
\caption{Simulated and measured values of the frequency difference $\Delta f_\textrm{max}$ in MHz between an empty sample cell (i.e. vacuum above the sample surface) and being fully covered in neon.}
\label{tab:tab1}
\end{table}

In order to obtain the model used in Fig. \ref{fig:fig2}b of the main text, we perform 2D cross-section simulations of the static capacitances between the center pin electrodes and ground with varying amounts of neon using Ansys Maxwell.
By utilizing known formulas for the effective capacitance of the common and differential modes of a coplanar stripline (see Ref. \cite{koolstra_trapping_2019}), we are able to translate between frequency shift and neon thickness.
We note that the galvanic connection of the common mode to the lumped on-chip filter is not accounted for in our modeling and results in some error.

The frequency shifts are scaled to the maximum observed frequency shift for the corresponding phase of neon.
Dielectric constants of $\epsilon^l_r=1.194$ and $\epsilon^s_r=1.244$ are used for the liquid and solid phases, respectively \cite{cole_image-potential-induced_1969}.
Numerical values of both the simulated and measured maximum frequency shifts are given in Table. \ref{tab:tab1}.

\section{\label{app:participation_simulations} Participation ratio simulations}

\begin{figure}[b]
\includegraphics{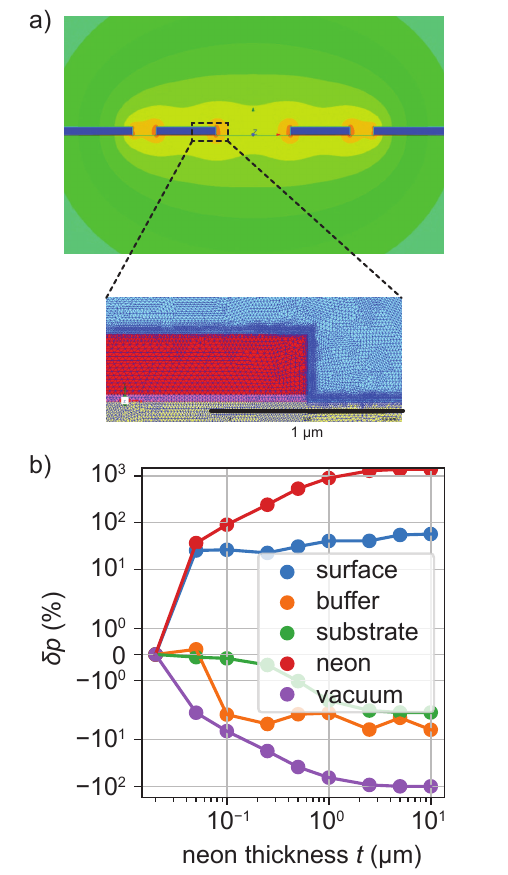}
\caption{\label{fig:sfig3} (a) Cross section of 2D finite element Ansys Maxwell simulation. The dielectric regions are chosen as the bulk sapphire ($\varepsilon_\textrm{sapphire} = 10$), the 40 nm thick layer of CeO$_2$ ($\varepsilon_{\textrm{CeO}_2} = 26$), a 10 nm thick surface layer (informed by AFM images that reveal significant surface roughness) that encompasses both the metal-air and substrate-air interfaces ($\varepsilon_\textrm{surf} = 10$), and the neon film. (b) Percent change in the participation ratio $\delta p = \big(p_i(t)-p_i(t_0)\big)/p_i(t_0)$ of each dielectric volume relative to a reference simulation with a uniform coverage of $t_0 = 20$ nm as a function of neon thickness $t$. }
\end{figure}

The participation ratio calculations for the loss estimate in Fig. \ref{fig:fig4} follow a simplified version of the methodology presented in Ref. \cite{wang_surface_2015}.
Typically, 3D simulations are used to extract the fields over a wide range of length scales for aperiodic and lumped-element geometries while 2D simulations are used to rescale divergent electric fields at structure edges and faces.
This allows for the capture of long-distance field variation due to device-scale boundary conditions while still accounting for electric field variation on dielectric boundaries at the nanometer scale.
There is a complex interface bandaging and rescaling method that links the two and three dimensional cases consistently and prevents double-counting. 
The differential mode of our coplanar stripline (CPS) structure can be accurately modeled with 2D simulations alone.
Since the differential mode leakage is limited beyond the node of the CPS, the CPS structure is entirely uniform in its cross section.
The ground plane is also nearby, so electric fields decay in magnitude radially with a constant on the order of the feature size, so the modal volume is very well represented even when only sampling within one or two orders of magnitude of the feature size.
This is in contrast to structures such as transmon capacitor pads, whose sizes are comparable to the substrate thickness, and n-polar coplanar waveguides with distant grounds such as those measured in cavities.
As a result, there is no bandaging or rescaling required since the 2D simulation samples the field well and more finely than a 3D simulation regardless.
If our resonators were of a higher quality factor, it would be necessary to follow other examples in dividing our surfaces and interfaces into more specific sections, but at this level it is appropriate to bound significant loss within regions with very small participation.
Finally, we note that our analysis provides a framework and quantitative estimates for determining how integrating condensed cryogenic films can provide insight towards dielectric loss mechanisms in state-of-the-art superconducting devices.

\bibliography{references_ybco}

\end{document}